\documentclass[aip,reprint, 10pt]{revtex4-1}

\usepackage{graphicx}
\usepackage{siunitx}

\usepackage{lineno}
\usepackage{amsmath}
\usepackage{amssymb}

\usepackage{mathtools}
\usepackage{enumitem}
\usepackage{xr-hyper}
\usepackage{hyperref}
        \hypersetup{ % for arXiv to compile correctly
           breaklinks=true,   % splits links across lines
           colorlinks=false,   % displays links as colored text instead of blocks
           pdfusetitle=true,  % \title and \author values into pdf metadata
        }

\usepackage[T1]{fontenc}
\usepackage[utf8]{inputenc}

\begin{document}

\title{Continuously tunable coherent pulse generation in semiconductor lasers}

\author{Urban Senica} 
 \thanks{Authors to whom correspondence should be addressed: usenica@phys.ethz.ch, scalari@phys.ethz.ch}
 \affiliation{Quantum Optoelectronics Group, Institute for Quantum Electronics, ETH Z{\"u}rich, 8093 Z{\"u}rich, Switzerland}
\author{Michael A. Schreiber} 
\affiliation{TUM School of Computation, Information and Technology, Technical University of Munich (TUM), 85748 Garching, Germany}
\author{Paolo Micheletti} 
\author{Mattias Beck}
 \affiliation{Quantum Optoelectronics Group, Institute for Quantum Electronics, ETH Z{\"u}rich, 8093 Z{\"u}rich, Switzerland}
\author{Christian Jirauschek} 
\affiliation{TUM School of Computation, Information and Technology, Technical University of Munich (TUM), 85748 Garching, Germany}
\author{J\'er\^ome Faist}
\author{Giacomo Scalari}
\thanks{Authors to whom correspondence should be addressed: usenica@phys.ethz.ch, scalari@phys.ethz.ch}
 \affiliation{Quantum Optoelectronics Group, Institute for Quantum Electronics, ETH Z{\"u}rich, 8093 Z{\"u}rich, Switzerland}

\date{\today}

\begin{abstract}
In a laser, the control of its spectral emission depends on the physical dimensions of the optical resonator, limiting it to a set of discrete cavity modes at specific frequencies. Here, we overcome this fundamental limit by demonstrating a monolithic semiconductor laser with a continuously tunable repetition rate from 4 up to 16 GHz, by employing a microwave driving signal that induces a spatiotemporal gain modulation along the entire laser cavity, generating intracavity mode-locked pulses with a continuously tunable group velocity. At the output, frequency combs with continuously tunable mode spacings are generated in the frequency domain, and coherent pulse trains with continuously tunable repetition rates are generated in the time domain. Our results pave the way for fully tunable chip-scale lasers and frequency combs, advantageous for use in a diverse variety of fields, from fundamental studies to applications such as high-resolution and dual-comb spectroscopy.

\end{abstract}

\maketitle

\section*{Introduction}
Ever since the first demonstrations of the maser by Schawlow and Townes \cite{SchawlowTownesMasers_PR1958} and of the optical laser by Maiman in 1960 \cite{Maiman_Nature1960}, a resonant optical cavity has been a crucial component of a given laser system, providing a set of cavity resonances which can be excited and amplified in combination with a gain medium. The possibility to excite one or many cavity modes is intrinsically related to the laser gain medium bandwidth, the physical dimensions of the cavity and the reflection coefficients of its output couplers, which have a fundamental role in the equations governing the laser dynamics \cite{siegman1986lasers}. In the simplest configuration of a Fabry-Pérot laser cavity with two partially reflective mirrors \cite{ismail2016fabry}, the frequency spacing between neighboring laser modes, or \textit{repetition rate} ($f_{\mathrm{rep}}$), is determined by the cavity resonances and is inversely proportional to the cavity length $L$ through the relation $f_{\mathrm{rep}}=\frac{c}{2n_{\mathrm{g}}L}$, where $c$ is the speed of light in vacuum and $n_{\mathrm{g}}$ is the group refractive index of the optical mode propagating within the laser medium \cite{siegman1986lasers}. 

Especially interesting is the case when the laser modes are locked through a nonlinear optical process, e.g., four-wave mixing, generating a spectrum with equidistant modes. These can be described by the relation $f_n=f_{\mathrm{ceo}}+n\times f_{\mathrm{rep}}$, where the carrier-envelope offset frequency $f_{\mathrm{ceo}}$ is related to the chromatic dispersion of the optical cavity. In the case where both $f_{\mathrm{ceo}}$ and $f_{\mathrm{rep}}$ are stabilized and locked to a referenced standard, a frequency comb is generated\cite{udem2002optical, diddams2020optical}, which can be described as a set of perfectly equidistant and phase-locked lasing modes in the frequency domain, corresponding to a perfectly periodic output waveform envelope in the time domain. Depending on the phases of the individual modes, the laser output in the time domain will consist of a train of (ultrashort) optical pulses (AM combs)\cite{keller2021ultrafast} or of a more complex periodic waveform (FM combs) \cite{hugi2012mid, burghoff2020unraveling,opavcak2019theory, senica2023frequency}. 

The repetition rate $f_{\mathrm{rep}}$ of a laser can be tuned by modifying the cavity resonances, either by changing the cavity length $L$ or the material permittivity $\varepsilon$.  With harmonic mode-locking, integer harmonics of the fundamental repetition rate can be excited as well using passive \cite{komarov2006passive} and active techniques\cite{becker1972harmonic}, or by generating multi-soliton states (Turing rolls)\cite{herr2014temporal}. However, these methods provide only a discrete change of the repetition rate. In a monolithic configuration, where the cavity dimensions are constant, $f_{\mathrm{rep}}$ can be continuously changed by a few percent at most \cite{Guo_MArandi_Science_2024}.

\subsection*{Frequency-tunable active mode-locking}
\begin{figure*}[tb]
\centering
\includegraphics[width=0.6\linewidth]{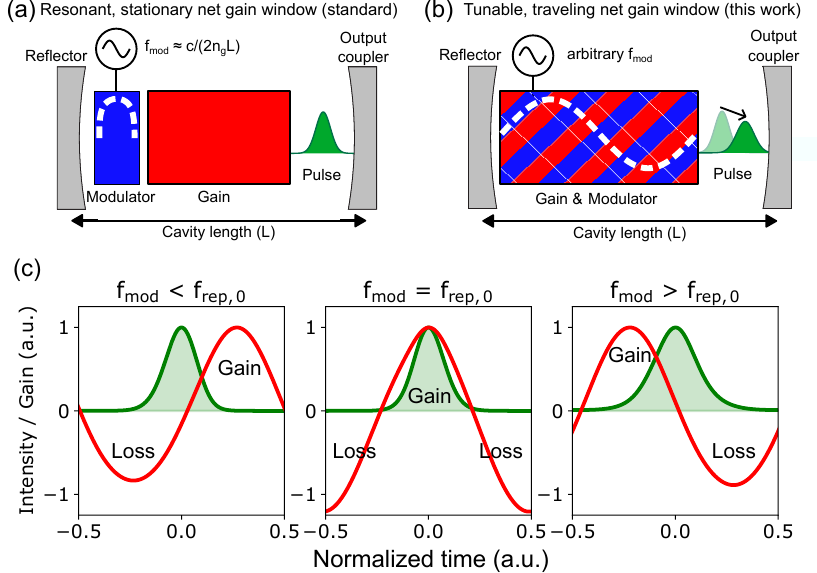}
\caption{\textbf{Overview of the general operating principle}. \textbf{(a)} In standard active mode-locking, the spatially confined modulator is synchronized to the cavity repetition rate, opening a stationary net gain window. \textbf{(b)} In our work, tunable active mode-locking is enabled by a spatiotemporal gain modulation along the entire lasing cavity. The resulting standing modulation wave (white dashed line) has a phase velocity that is proportional to the modulation frequency, generating a tunable traveling net gain window. \textbf{(c)} Results of our numerical model in three different regimes in a co-propagating frame (averaged over the forward propagation cycle). Any shift between the peak of the optical pulse and the modulation wave results in a gain/loss gradient throughout the optical pulse, effectively speeding it up or slowing it down, such that the new repetition rate is synchronized to the external modulation frequency.}
\label{fig:Concept}
\end{figure*}

In standard active mode-locking, a modulator element is placed within the laser cavity, as illustrated in Fig. \ref{fig:Concept}(a). Its role is to modulate the optical gain (or losses), such that a short net gain time window opens, allowing for the generation of short pulses. The modulator element is typically separated from the gain element and is spatially confined to a small fraction of the laser cavity. In this case, the modulator frequency needs to be resonant with the cavity repetition rate, as a spatially stationary net gain window is generated. Detuning the modulator frequency by a significant fraction of the repetition rate hinders the formation of short pulses, as the modulator element cannot affect the light propagation throughout most of the laser cavity, which would fall out of sync with the detuned net gain window.

In this work, we propose a novel regime, where the repetition rate of a mode-locked semiconductor laser can be tuned continuously and significantly both above and below the fundamental frequency defined by the cavity length. As sketched in Fig. \ref{fig:Concept}(b), this is enabled by a spatiotemporal gain modulation along the entire laser device via an external microwave signal that modulates the laser bias. In this configuration, the modulator can induce a spatially extended gain/loss wave, resulting in a traveling net gain window that affects the optical pulse propagation along the entire device length. It is important to note that the distributed modulator element should generate a standing gain/loss modulation wave. This way, its phase velocity, i.e., the propagation velocity of the traveling net gain window, is proportional to the modulation frequency, which is crucial for generating tunable optical pulses. In a real device, this can be implemented by injecting microwaves at one end of the low-loss waveguide and imposing reflective boundary conditions, such that the interference between counterpropagating microwaves generates a (quasi-) standing wave.

To understand the basic properties and stability of such a system, we developed an analytical model by expanding the Haus-type master equation framework presented in Refs. \cite{haus1975theoryC,KaertnerPhysRevLett1999}. These models describe the optical pulse by its roundtrip-averaged envelope, giving Gaussian pulse solutions for arbitrary values of detuning from the roundtrip frequency. While the detailed derivation of our model is in the Supplementary Material, we present here the main expression describing the evolution of the electric field $E^{\pm}$ in a co-propagating frame, neglecting chromatic dispersion:

% equations from TUM
\begin{equation}
\pm\partial_{x}E^{\pm}=\delta^{\pm}\partial_{\tau}E^{\pm}+\frac{1}%
{2}\left(  g^{\pm}+g_{\omega
}\partial_{\tau}^{2}-a\right)  E^{\pm},
\label{eq:analytical_eq}%
\end{equation}
with the detuning parameter $\delta^{\pm}=\left(  v_{\mathrm{m}}^{\pm}\right)
^{-1}-v_{\mathrm{g,0}}^{-1}$ (defined as the difference between the inverse optical group velocities in the modulated and unmodulated cavity), the saturated time-modulated gain $g^{\pm}$, the spectral gain bandwidth in parabolic approximation $g_{\omega
}$ and the loss coefficient $a$. By assuming a Gaussian pulse shape, a stationary analytical solution can be obtained, which relates the detuning $\delta^{\pm}$ with the gain slope $g_{1}^{\pm}$ and the FWHM pulse duration $T^{\pm}$:

\begin{equation}
\delta^{\pm}=g_{1}^{\pm}\left(  T^{\pm}\right)  ^{2}/\left(  8\ln2\right).
\label{eq:analytical_pulse}%
\end{equation}

We can now develop an intuitive understanding of the tunable mode-locking mechanism. In Fig. \ref{fig:Concept}(c), we plot the numerical model solutions, averaged over the forward propagation cycle, for three different typical regimes in the co-propagating frame. These have been obtained with a semiclassical Maxwell-density matrix formalism which features a generalized multilevel Hamiltonian, containing light-matter interaction as well as tunneling across thick barriers (see Supplementary Material, equations 1-3). In order to reduce the numerical load, we invoke the rotating wave approximation. The RF modulation is modeled by solving the transmission line equations across the waveguide and considering the bias dependence of the active region Hamiltonian and Lindbladian \cite{jirauschek2023theory}. 

Since the period of the modulation wave is determined by the external modulation frequency $f_{\mathrm{mod}}$, the traveling net gain window completes a full roundtrip in time $1/f_{\mathrm{mod}}$. When the modulation frequency matches the natural repetition rate ($f_{\mathrm{mod}}=f_{\mathrm{rep,0}}$), the group velocity of the optical pulse and the phase velocity of the modulation wave are also matched naturally. The peaks of the optical pulse and the gain/loss modulation wave are aligned (synchronized), converging to the standard active-mode locking case. However, when the modulation frequency is positively detuned ($f_{\mathrm{mod}}>f_{\mathrm{rep,0}}$), an optical pulse initially propagating with the natural group velocity will be slower than the traveling gain/loss window, thus experiencing a gain/loss gradient, i.e., more gain at its leading than its trailing edge, which will experience loss. For sufficiently large modulation amplitudes, this will continuously deform the propagating pulse, effectively increasing its group velocity and matching it to the increased velocity of the traveling net gain window. The opposite effect will happen for modulation frequencies below the natural repetition rate ($f_{\mathrm{mod}}<f_{\mathrm{rep,0}}$), i.e., the pulse will be initially faster than the gain/loss wave and effectively slowed down. These different regimes are clearly visible in Fig. \ref{fig:Concept}(c), where the optical pulse is either aligned with, ahead of, or lagging behind the peak of the modulation wave. In all the cases, the average group velocity of the intracavity optical waveform envelope will be adjusted such that the roundtrip time is perfectly synchronized to the modulation frequency. 

\begin{figure*}[tbp]
\centering
\includegraphics[width=0.9\linewidth]{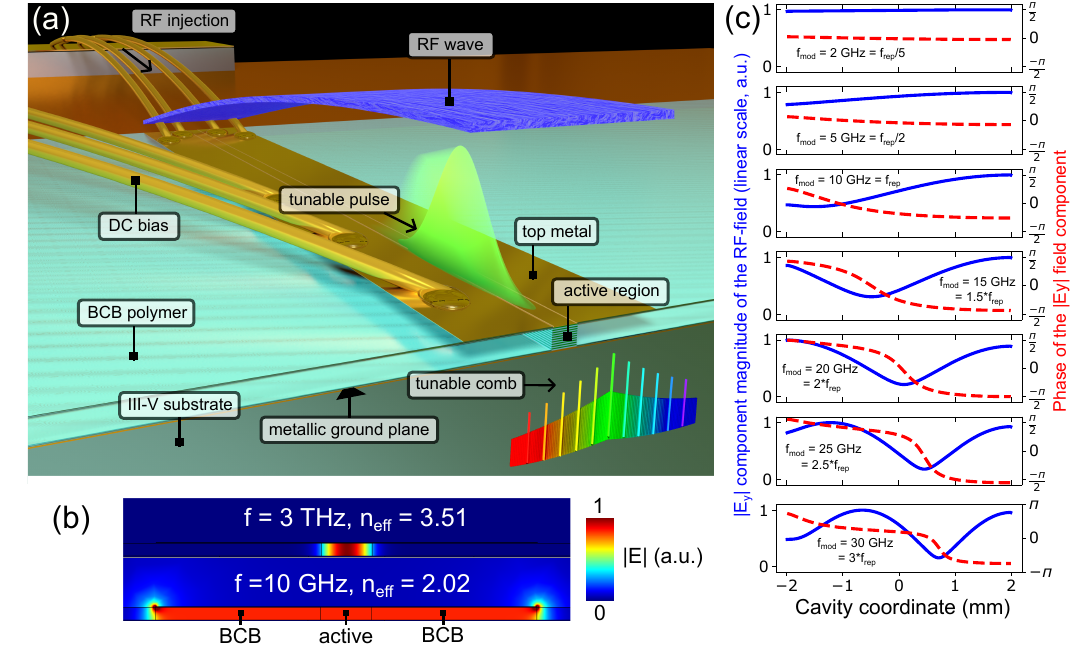}
\caption{\textbf{Device geometry and basic properties.} \textbf{(a)} Illustration of the experimental platform, where microwaves are injected into a planarized waveguide with a semiconductor laser core (THz QCL). A radio-frequency (RF) wave (blue) is formed along the entire lasing cavity, generating intracavity mode-locked pulses (green) with a tunable group velocity. \textbf{(b)} Simulated 2D waveguide cross-section electric field intensity profiles for the propagating electromagnetic mode at the lasing frequency (3 THz, confined to the active core) and at microwave frequencies (10 GHz, spread across the planarized waveguide). \textbf{(c)} 3D simulation results of the RF field distribution along the center of a 4 mm long active waveguide, when injected from the back facet (left side), forming a frequency-dependent wave pattern. The amplitude and phase are shown in blue and red, respectively.}
\label{fig:Overview}
\end{figure*}

By comparing the simulation results to equation~\ref{eq:analytical_pulse} obtained by the analytical model, it can be shown that the pulse pulling due to the gain slope explains the continuous tunability (see Supplementary material, Table S1). This way, a coherent pulse train with an arbitrary repetition rate can be synthesized by simply tuning the microwave driving signal over a wide range of frequencies, regardless of the amount of detuning to the natural cavity repetition rate. In the frequency domain, this generates frequency combs with a continuously tunable mode spacing. The crucial aspect is that the tunable mode-locked laser is no longer limited to lase on the usual longitudinal cavity modes imposed by the device dimensions. Instead, virtually any discrete lasing frequency within the gain bandwidth can be excited, as the frequencies of the lasing lines are simply a Fourier transform of the continuously tunable periodic pulse train at the output.

\subsection*{Experimental platform}
For the experimental demonstration, we use a broadband planarized terahertz quantum cascade laser (THz QCL) \cite{senica2022planarized}. In general, QCLs are electrically pumped semiconductor lasers based on intersubband transitions \cite{faist1994quantum}, allowing for narrow- and broadband emission at relatively long wavelengths in the mid-infrared (mid-IR) \cite{yao2012mid}  and terahertz (THz) \cite{kohler2002terahertz} regions of the electromagnetic spectrum. This gain medium features ultrafast dynamics and large optical nonlinearities that enable the locking of broadband emission spectra into coherent frequency combs \cite{hugi2012mid, BurghoffNatPhot2014, faist2016quantum}, useful for applications in spectroscopy and sensing \cite{picque2019frequency}. In our case, we employ a homogeneous broadband active region centered around 3 THz \cite{forrer2020photon} that allows for broadband and frequency comb operation with bandwidths larger than 1 THz. 

As illustrated in Fig. \ref{fig:Overview}(a), the planarized THz QCL device consists of an active material waveguide embedded in the low-loss polymer Benzocyclobutene (BCB) and sandwiched between a top and bottom metallic layer, vertically confining the optical and microwave modes and acting as electrical contacts. This configuration is advantageous in several aspects, such as low waveguide losses, efficient heat dissipation, and excellent microwave properties. In particular, as shown in the waveguide 2D cross-section mode simulations in Fig. \ref{fig:Overview}(b), while the optical mode is confined within the central active material waveguide, the microwave mode spreads across the entire extended top and bottom metallization. With a reduced spatial overlap with the lossy central active waveguide (arising from its finite vertical conductivity), the planarized waveguide acts as a low-loss, low-impedance microstrip transmission line for microwaves (these aspects were explained in greater detail in Ref. \cite{senica2022planarized}). As microwaves are injected via a set of short, straight bonding wires at the back waveguide facet, they propagate and reflect at the front waveguide facet, forming a frequency-dependent wave pattern along the entire laser cavity. As shown in 3D numerical simulation results in Fig. \ref{fig:Overview}(c), the radio-frequency (RF) field distribution within the waveguide displays a maximum at the front waveguide facet. For low frequencies ($f_{\mathrm{mod}}\ll f_{\mathrm{rep,0}}$), the modulation is in a quasi-static regime with a nearly uniform field intensity oscillating in phase.  With increasing frequency, an increasing number of minima and maxima form along the cavity, and the wave distributions resemble an oscillating cosine function in a time-domain field animation. These properties are crucial, as they allow for an efficient spatiotemporal gain modulation for an arbitrary microwave modulation frequency, regardless of any mismatch between the waveguiding properties at microwave and terahertz frequencies, such as their effective mode indices and propagation velocities. This differs from many other integrated microwave photonics devices, where more elaborate phase-matching schemes are required for an efficient co-propagation of optical and microwave modes at specific frequencies \cite{marpaung2019integrated}.

\section*{Results}
\subsection*{Coherent measurements and simulations}
Experimentally, we investigated a 6 mm long planarized ridge waveguide, where the central active waveguide is 40 \SI{}{\micro\metre} wide, and the extended top metallization has a width of 300 \SI{}{\micro\metre}.  The device is measured at a heat sink temperature of 40 K, and the natural repetition rate is $f_{\mathrm{rep,0}}=6.61$ GHz.

\begin{figure*}[tbp]
\centering
\includegraphics[width=0.8\linewidth]{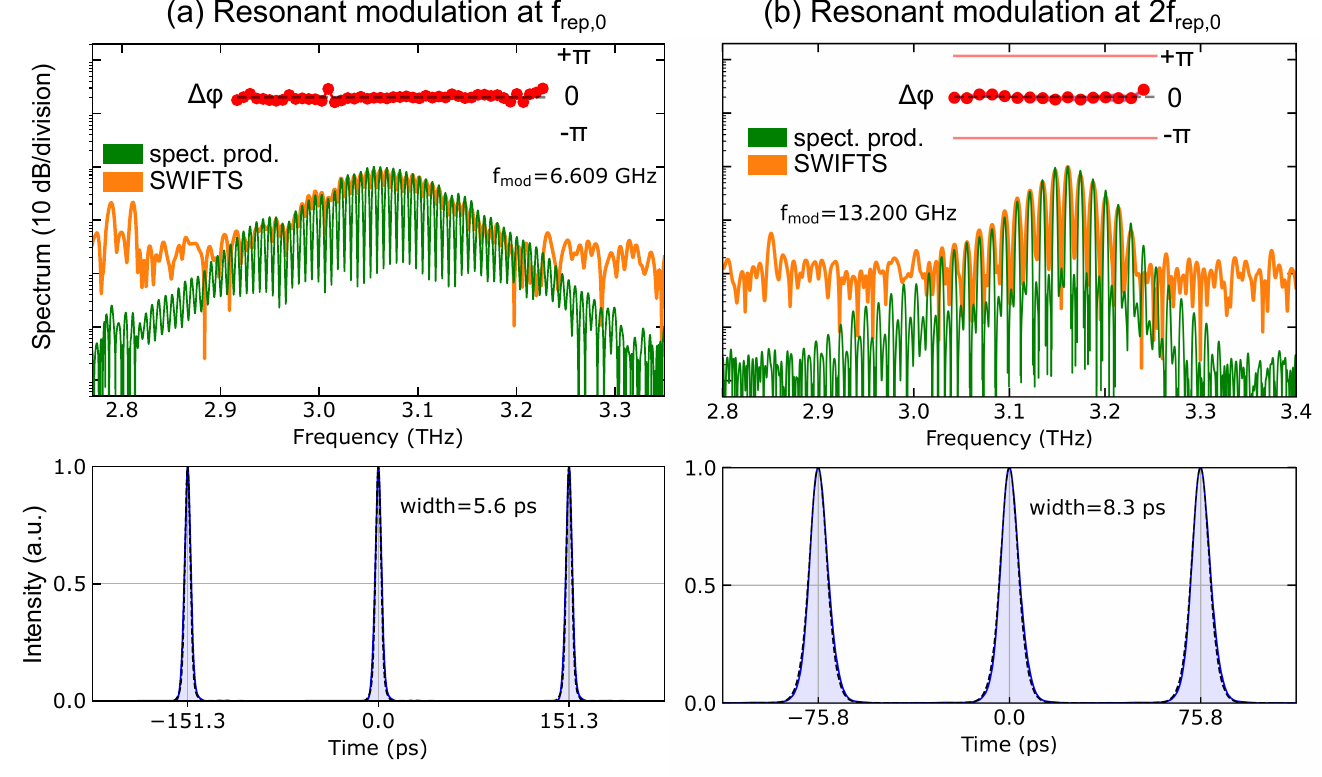}
\caption{\textbf{Experimental SWIFT spectroscopy measurement results under resonant conditions.} Microwave modulation at a frequency of \textbf{(a)} $f_{\mathrm{rep,0}}$ and \textbf{(b)} $2f_{\mathrm{rep,0}}$ results in the broadest spectral bandwidths and the shortest pulses, as the system operates in the standard active mode-locking regime, i.e., without any significant detuning between the modulation and natural repetition frequencies.}
\label{fig:SWIFTS_resonant}
\end{figure*}

We performed a modulation frequency sweep study, where the DC laser bias is kept close to the laser threshold and a strong microwave signal (typically +30 dBm at the source) is injected at the back laser facet. To assess the coherence of the measured emission spectra and to reconstruct the time-domain output intensity, we performed Shifted Wave Interference Fourier Transform (SWIFT) spectroscopy \cite{BurghoffNatPhot2014, han2020sensitivity}. Using this method, the optical beatnote generated between adjacent lasing modes is measured with a fast detector (bandwidth $>$ 30 GHz) after passing through a Fourier Transform Interferometer (FTIR) \cite{berthomieu2009fourier}, which acts as a tunable frequency-selective filter.
Under standard (resonant) driving conditions, the microwave injection frequency is close (to within a few MHz) to the natural repetition rate, and resonant modulation sidebands are generated which can injection-lock the longitudinal cavity modes, stabilizing the periodic laser output and broadening the emission spectrum \cite{hillbrand2019coherent, schneider2021controlling}. These experimental results are shown in Fig. \ref{fig:SWIFTS_resonant} for the case of fundamental and second harmonic mode-locking, where the spectrum product measured directly with a slow detector (DTGS) and the SWIFT spectrum obtained via the optical beatnote using a fast detector (Schottky diode) have an excellent agreement. The latter allows us to obtain the intermodal phase differences and reconstruct the time-domain output profile, producing spectral bandwidths up to 500 GHz and short pulses down to 5.6 ps under resonant modulation conditions. The exact resonant frequency was found experimentally by biasing the laser at the threshold and performing a fine sweep of the modulation frequency within $\pm$10 MHz around the measured free-running repetition rate.

\begin{figure*}[tbp]
\centering
\includegraphics[width=0.9\linewidth]{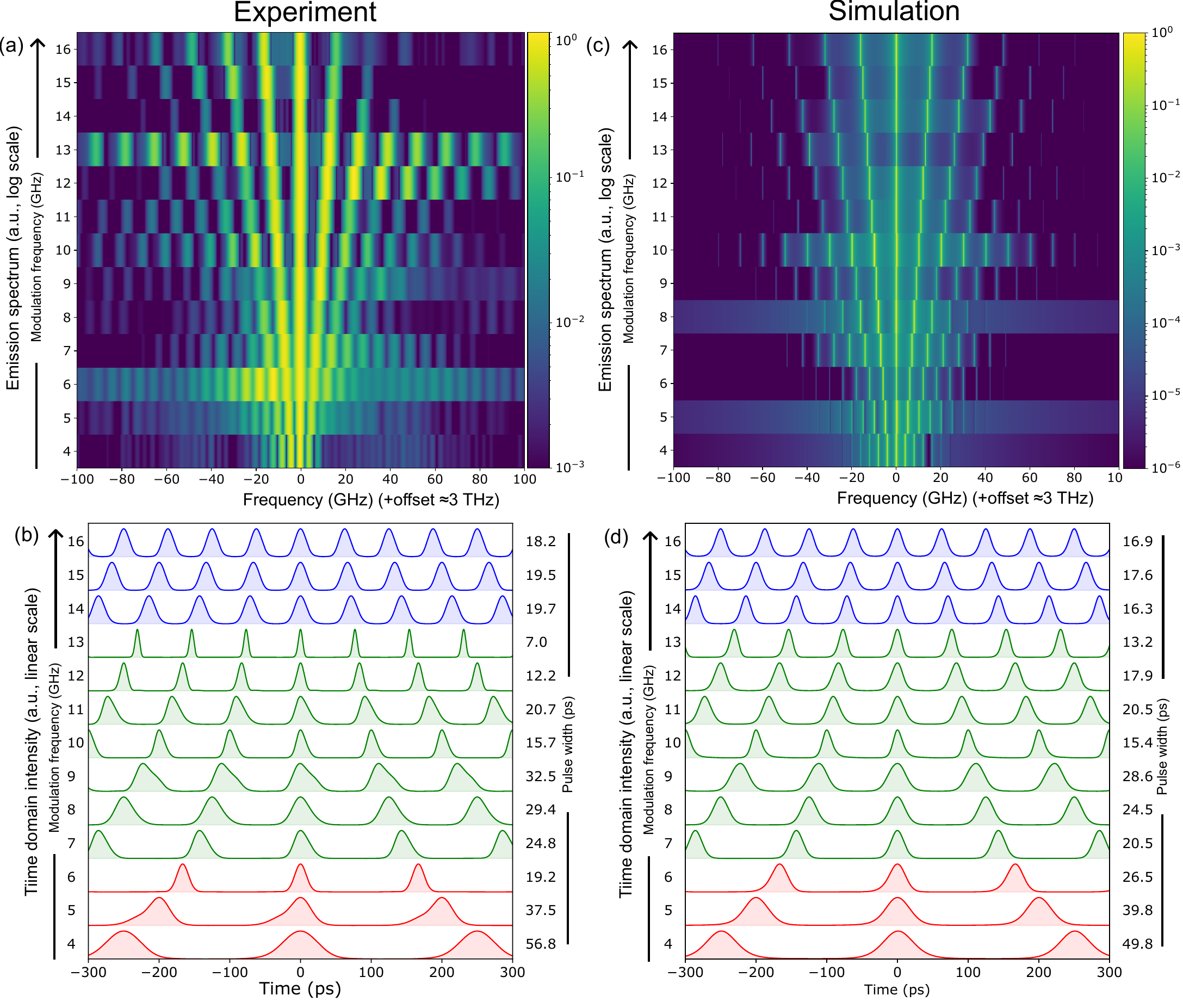}
\caption{\textbf{Spectral and time-domain measurement and numerical simulation results}.  \textbf{(a)} Measured emission spectra for modulation frequencies between 4 and 16 GHz, where the spectra are aligned and zoomed around 200 GHz of the strongest central mode. \textbf{(b)} The reconstructed time-domain intensity profiles reveal that a set of coherent pulse trains is generated, where the tunable repetition rate matches the microwave modulation frequency. The simulated emission spectra \textbf{(c)} and time-domain intensity profiles \textbf{(d)} show excellent agreement with experimental results, displaying continuously tunable mode spacings and repetition rates, synchronized to the microwave modulation frequency between 4 and 16 GHz.}
\label{fig:Simulation}
\end{figure*}

In our experimental configuration, we can drive the laser system beyond the conventional regime of exciting longitudinal cavity modes and instead generate tunable emission spectra where the mode spacing matches the microwave modulation frequency. This novel regime is accessible regardless of the amount of detuning to the natural repetition rate of the laser cavity, generating entirely new lasing modes at different frequencies. Remarkably, such tunable spectra were obtained even for extreme values of frequency detuning, ranging from nearly $f_{\mathrm{rep,0}}/2$ to well above $2f_{\mathrm{rep,0}}$. The measurements were performed across a relatively large modulation frequency range between 4 and 16 GHz with a step of 1 GHz, corresponding to a relative repetition frequency tuning range of 400\%. In Fig. \ref{fig:Simulation}(a), we show a 2D colormap of the measured spectra as a function of modulation frequency, aligned and zoomed within 200 GHz of the strongest central mode (the full SWIFTS results for each individual measurement are in the Supplementary Material, Fig. S2).  The tunable mode spacing is clearly visible as opening lines on the spectral map. As shown in Fig. \ref{fig:Simulation}(b), a set of coherent pulse trains with a tunable repetition time is generated in the time domain. The individual pulse durations of these coherent frequency comb states are shorter for smaller frequency detunings from the natural repetition rate (or its integer multiples) due to a resonant effect, and also for higher modulation frequencies due to shorter modulation period times. The slight asymmetry and its orientation in some of the pulse shapes ($f_{\mathrm{mod}} = 5,9,11$ GHz) is a signature of the pulse formation mechanism envisioned in Ref.\cite{KaertnerPhysRevLett1999}.

Such detailed features of the detuned pulse solutions can be recovered in numerical simulations of the modulated laser cavity. Our dynamic model is based on the semiclassical Maxwell-density matrix formalism which features a generalized multilevel Hamiltonian (see above and Supplementary Material, equations 1-3). The resulting emission spectra and time-domain intensity output are displayed in Fig. \ref{fig:Simulation}(c, d), showing excellent agreement with experimental data and reproducing both the tunable mode spacing and repetition rate of the coherent pulse train. The linewidth of the resulting comb lines is far below the numerical resolution limit, which is 1 MHz for our Fourier transformation of almost 7000 round trips. %The slight pulse asymmetry and its orientation, although less pronounced, also agree with experimental observations.

To prove that the simulated optical fields are mode-locked, we evaluate the power and phase noise quantifiers $M_\mathrm{\Delta P}$ and $M_\mathrm{\Delta \Phi}$ from Ref.~\cite{silvestri2020coherent}. This analysis shows, that all injection frequencies in Fig.~\ref{fig:Simulation} fulfill the mode-locking conditions, i.e. $M_\mathrm{\Delta P} < 0.005\,P_\mathrm{av}$ and $M_\mathrm{\Delta \Phi} < 0.02$, where $P_\mathrm{av}$ denotes the roundtrip-averaged optical power.

\begin{figure*}[tbp]
\centering
\includegraphics[width=0.9\linewidth]{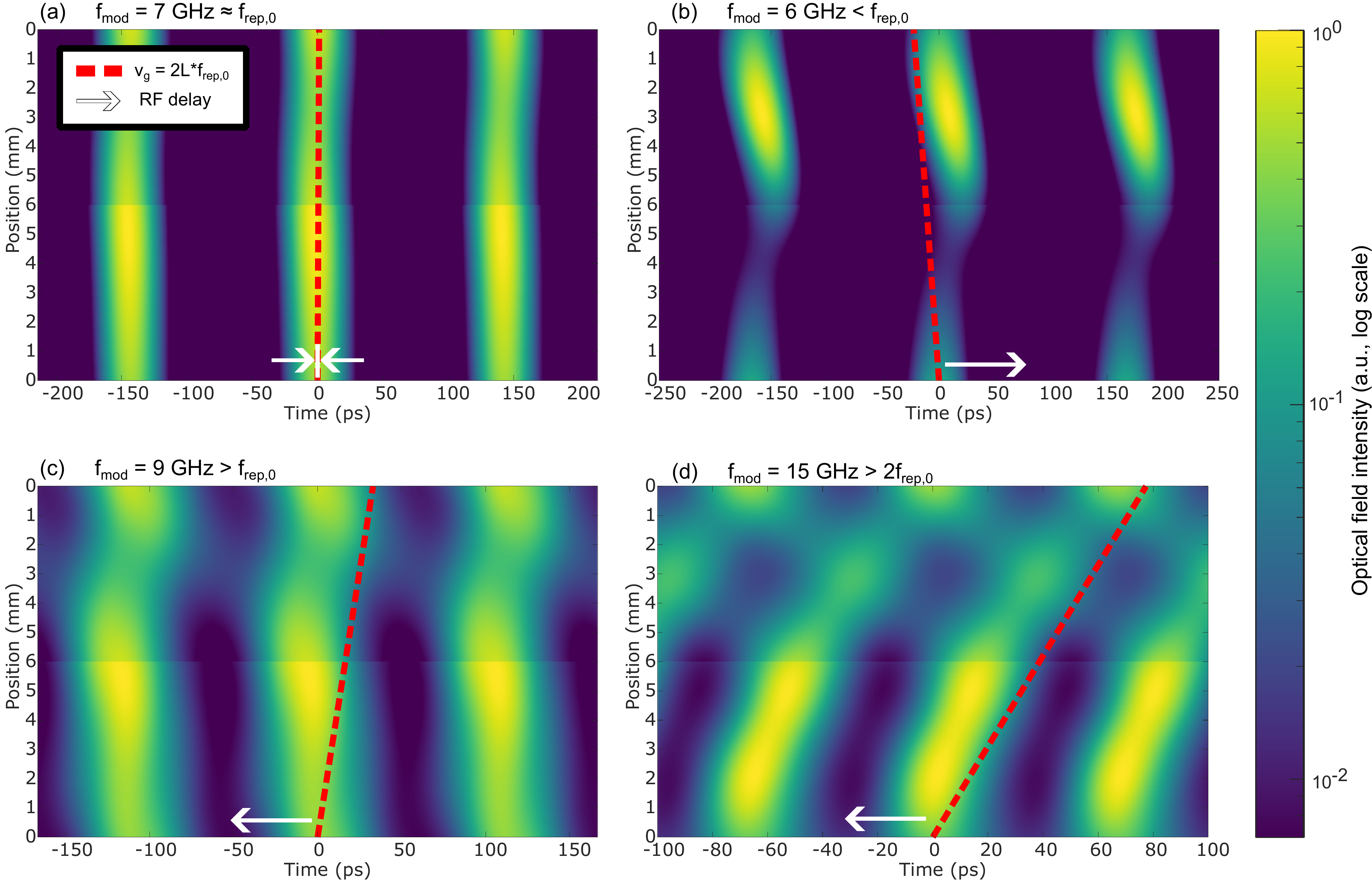}
\caption{\textbf{Simulated intracavity waveform dynamics.} 
Comparison of the intracavity optical field intensities for different modulation frequencies of \textbf{(a)} 7 GHz $\approx f_{\mathrm{rep,0}}$, \textbf{(b)} 6 GHz $< f_{\mathrm{rep,0}}$, \textbf{(c)} 9 GHz $> f_{\mathrm{rep,0}}$, and \textbf{(d)} 15 GHz $> 2f_{\mathrm{rep,0}}$, showing three roundtrip times of each periodic waveform in the unfolded cavity representation in the co-propagating reference frame. Microwaves are injected at the cavity coordinate of 0~mm. While the dashed red lines show the reference optical group velocity under free-running or resonant modulation conditions ($f_{\mathrm{mod}}=f_{\mathrm{rep,0}}$), the white arrows indicate the delay (phase shift) between the maxima of the modulation RF wave and the optical pulse, showing the direction of the resulting pulse pulling effect (see also Fig. \ref{fig:Concept}(c)). Any deviation from the straight vertical direction in the individual panels indicates a local variation of the optical pulse group velocity. At the central horizontal line (cavity coordinate of 6 mm), the optical field intensity is reduced due to the finite reflectivity at the waveguide facet. The time-domain intensity profiles from Fig.~\ref{fig:Simulation}~(d) are taken at this point.
}
\label{fig:IntracavityDynamics}
\end{figure*}

\subsection*{Intracavity waveform dynamics}
In contrast to the analytical model, which only gives roundtrip averaged solutions, we can employ numerical simulations of the driven laser cavity to study the intracavity waveform dynamics. In Fig. \ref{fig:IntracavityDynamics}, we plot the intracavity optical field intensities in a 2D colormap using a co-propagating reference frame determined by the modulation frequency $f_{\mathrm{mod}}$.  The data are shown in logarithmic scale, as the waveform amplitudes can change significantly within a single roundtrip time. For each case, three roundtrip times are displayed in the unfolded cavity representation, i.e., the time range is set between (-1.5/$f_{\mathrm{mod}}$, 1.5/$f_{\mathrm{mod}}$), and we plot the counter-propagating waves separately in a stitched/mirrored fashion. In this view, a vertical line connecting the bottom and top edge of the plot represents propagation at the new roundtrip-averaged group velocity of $v_{\mathrm{g}}=2Lf_{\mathrm{mod}}$, and a pulse completes a single roundtrip along the unfolded cavity by traversing the entire plot range in the vertical direction. The red dashed line marks the reference group velocity, representing the pulse propagation in the resonant case ($f_{\mathrm{mod}}=f_{\mathrm{rep,0}}$). The white arrows indicate the roundtrip-averaged delay between the peak of the RF modulation wave and the optical pulse, shown already in Fig. \ref{fig:Concept}(c), marking the direction of the pulse pulling effect. In Fig. \ref{fig:IntracavityDynamics}(a), we first display the standard (resonant) active mode-locking case, where the optical pulse propagates at the natural group velocity without any significant perturbations, and there is no delay between the RF modulation wave and the optical pulse.

We now study the intracavity wave dynamics for three distinct regimes of detuning observed within our experimental range between 4-16 GHz. In panel (b), the modulation frequency of 6 GHz is below $f_{\mathrm{rep,0}}$, and the intracavity waveform features a significant change in its shape, amplitude, and local group velocity throughout the roundtrip while remaining on the right side of the reference red line, displaying a slowed-down pulse propagation. The white arrows indicate that the optical pulse is ahead of the RF modulation wave, experiencing a negative gain/loss slope, which slows down the pulse propagation. In panel (c), where $f_{\mathrm{mod}} = 9$ GHz $> f_{\mathrm{rep,0}}$, the average group velocity of the optical waveform is faster than the natural one, with visible changes (oscillations) of the local group velocity throughout the roundtrip time. In this case, the optical pulse lags behind the peak of the RF wave, experiencing a positive gain/loss slope, resulting in a speed-up of the optical pulse propagation. 

Another interesting regime is observed for modulation frequencies above $2f_{\mathrm{rep,0}}$, as shown in panel (d) for $f_{\mathrm{mod}}=15$ GHz, where the optical pulses seem to propagate along a diagonal instead of a vertical direction in the co-propagating reference frame. This indicates the formation of a double counter-propagating pulse structure, meaning that a single pulse only needs to travel half of the usual path for the waveform to complete a full signal period. This way, the propagation of each optical pulse itself is not much faster than the natural one (reference red line). The fact that several intracavity pulses form when the modulation frequency is above integer multiples of the natural repetition rate, akin to resonant harmonic mode-locking\cite{becker1972harmonic}, is an indicator that very high modulation frequencies should be possible without the requirement of extreme (superluminal) group velocities, as the waveform is instead split into several pulses. 

The diverse landscape of intracavity waveform dynamics showcased in the selected frequency examples is related to the specific spatiotemporal distribution of the modulation microwaves, which depends on the electrode/microwave waveguide geometry and its frequency-dependent properties, such as impedance, losses, reflectivity, and the resulting interference between counter-propagating microwaves, as shown already in Fig. \ref{fig:Overview}(c). This also influences the resulting output pulses in terms of their duration, shape, and asymmetry, as observed in experiments and simulations. In the Supplementary Material, we also added another version of the figure using real space-time coordinates (the lab reference frame) along with animated movies of the RF and optical fields, which can be used to study their dynamics as a function of space and time, see Figs. S3 and S4.

\subsection*{Comb properties and generalizations}

In our experiments, the highest obtainable modulation frequency of 19 GHz was limited by the upper-frequency cut-off of the microwave generator and amplifier equipment. Numerical simulation results suggest that the modulation frequency can in principle be very high, with a fundamental upper limit imposed by the laser gain bandwidth (Fig. S7). On the other hand, the lowest modulation frequency where coherent mode-locked pulses were still observed experimentally was at $f_{\mathrm{mod}}$ = 3.5 GHz, which is slightly above $f_{\mathrm{rep,0}}/2$ = 3.3 GHz. Our measurements indicate that with such a slow modulation, the laser operation shifts towards a gain-switched regime \cite{taschler2023short}, producing broadband, low-coherence emission \cite{CargioliAPLPhot2024}. Both the coherent and the gain-switched spectra obtained at the low-frequency modulation limit are shown in Fig. S8 of the Supplementary Material.

Additionally, we can evaluate several of the tunable comb properties with numerical simulations. As shown in Fig. S10, the comb is typically stabilized within around 1 \SI{}{\micro\second}, amounting to $\sim$$10^4$ completed cavity roundtrip times. 
A necessary condition for the formation of these coherent states is a sufficient modulation amplitude across the entire laser cavity. The tunable coherent pulse train does not form with low modulation amplitudes, or with large microwave propagation attenuation, where the microwaves are absorbed before reaching the front laser facet. Another requirement is that the laser is biased close to the lasing threshold, as this generates dynamic regions of both gain and loss. With an increasing DC laser bias, the laser regime first transitions to broader emission spectra, where individual sub-combs still feature the mode spacing following the microwave modulation frequency but are separated into several spectral portions with different $f_{\mathrm{ceo}}$ offset frequencies (see also Fig. S9). A further increase of the laser bias results in emission spectra with the natural mode spacing imposed by the cavity boundary conditions. This occurs due to an increasing gain saturation, which decreases the effective modulation amplitude, eventually hindering the formation of tunable spectra. A more detailed study of these aspects can be found in the Supplementary Material.

\section*{Conclusion}
We have demonstrated, both in experiments and simulations, how a strong spatiotemporal gain modulation along the entire semiconductor laser cavity can sculpt the intracavity waveforms into coherent pulse trains with continuously tunable repetition rates in the time domain, and frequency combs with continuously tunable mode spacings in the frequency domain. Within the cavity, both slow and fast light \cite{boyd2009slow, thevenaz2008slow} with a continuously tunable group velocity and broad spectral coverage can be generated via a pulse pulling effect. Besides discovering a novel regime that transcends the conventional concept of static cavity modes, this also paves the way for many interesting applications. In the frequency domain, the tunable mode spacing could be used for high-resolution and dual-comb spectroscopy to scan across absorption features without any spectral gaps. We explored this possibility by using a tapered waveguide device \cite{senica2023frequency} that generates tunable spectra with a stable central frequency, resulting in continuous gapless spectral scans (Fig. S15), as explained in more detail in the Supplementary Material. In the time domain, the coherent pulse train can be synchronized to an arbitrary external microwave frequency reference, acting as an electronically tunable delay line without any movable parts. 

These tunable states are robust, stable, reproducible, and fully determined by the microwave driving field and the laser bias. We have essentially demonstrated a functionality similar to tunable external cavity lasers \cite{mroziewicz2008external}, where the repetition rate is changed by modifying the cavity dimensions, but on a compact chip-scale device with no movable parts (more robust to mechanical vibrations) and with fully electronic control. This also enables a much faster frequency tuning, limited by the comb stabilization time on the order of 1 \SI{}{\micro\second}.
%mention also less sensitive to mechanical vibrations? Since mode spacing is fully determined by the RF source.
Compared to non-resonant electro-optic (EO) frequency combs \cite{parriaux2020electro} that also display wide repetition rate tunability, our device features a monolithic configuration and relatively wide spectral coverage that can be achieved without recurring to cascaded phase modulators and/or non-linear optical elements (fibers, waveguides) frequently employed to broaden the EO comb bandwidth.  

In conclusion, the crucial building blocks for such a widely tunable coherent pulse generation are an efficient spatiotemporal gain modulation across the entire laser cavity and fast gain that enables sculpting the intracavity waveform to be synchronized to the external microwave drive. 
While we demonstrated this concept with THz QCLs, this could in principle be expanded to other wavelength regions, by employing suitable modulation-optimized geometries and exploiting fast gain components also in interband laser devices \cite{MarzbanQwalktelecom2024} across the entire infrared and visible spectral region with a vast range of applications in sensing and telecommunications.

\section*{Acknowledgments}
U.S. and P.M. thank A. Forrer for his earlier work on the SWIFTS setup and analysis code. G.S. would like to acknowledge discussions with A. Dikopoltsev and A. Eichler. Funding from H2020 European Research Council Consolidator Grant (724344) (CHIC) and SNF project 200021-212735 are gratefully acknowledged.

% add citations from the Supplementary
\phantom{\cite{bozhkov2003semiconductor, jirauschek2019optoelectronic, risken1968self}}

\section*{References}\label{References}
\bibliography{arxiv}

\end{document}